\pdfoutput=1
\documentclass[%
 reprint,
 amsmath,amssymb,
 aps,prl
]{revtex4-1}

\usepackage{graphicx}
\usepackage{dcolumn}
\usepackage{bm}

\bmdefine{\boldi}{i}
\bmdefine{\boldj}{j}
\bmdefine{\boldS}{S}
\bmdefine{\boldq}{q}
\bmdefine{\boldzero}{0}
\bmdefine{\bolddelta}{\delta}

\begin{document}

\title{
Stabilizing Mechanism for Bose-Einstein Condensation of Interacting Magnons 
\\in Ferrimagnets and Ferromagnets}

\author{Naoya Arakawa}
\email{naoya.arakawa@sci.toho-u.ac.jp} 
\affiliation{
Department of Physics, Toho University, 
Funabashi, Chiba, 274-8510, Japan}

\date{\today}

\begin{abstract}
We propose a stabilizing mechanism for the Bose-Einstein condensation (BEC) of interacting magnons 
in ferrimagnets and ferromagnets. 
By studying the effects of the magnon-magnon interaction on the stability of 
the magnon BEC in a ferrimagnet and two ferromagnets, 
we show that 
the magnon BEC remains stable 
even in the presence of the magnon-magnon interaction 
in the ferrimagnet and ferromagnet with a sublattice structure, 
whereas it becomes unstable in the ferromagnet without a sublattice structure. 
This indicates that 
the existence of a sublattice structure is the key to stabilizing 
the BEC of interacting magnons, 
and the difference between the spin alignments of a ferrimagnet and a ferromagnet is irrelevant. 
Our result can resolve a contradiction between experiment and theory 
in the magnon BEC of yttrium iron garnet. 
Our theoretical framework may provide a starting point for understanding 
the physics of the magnon BEC including the interaction effects. 

\end{abstract}
\maketitle


Bose-Einstein condensation (BEC) 
has been extensively studied in 
various fields of physics. 
The BEC is a macroscopic occupation of 
the lowest-energy state for bosons~\cite{BEC-text}. 
This phenomenon was theoretically predicted 
in a gas of noninteracting bosons~\cite{Einstein}, 
and then it was experimentally observed in dilute atomic gases~\cite{BEC-exp1,BEC-exp2,BEC-exp3}. 
This observation opened up research of the BEC 
in atomic physics~\cite{BEC-text}. 
Since the concept of the BEC is applicable to quasiparticles 
that obey Bose statistics, 
research of the BEC has been expanded, 
and it covers condensed-matter physics, 
nuclear physics, 
and optical physics. 

There is a critical problem with the magnon BEC. 
The magnon BEC was experimentally observed in yttrium iron garnet (YIG), 
a three-dimensional ferrimagnet~\cite{magBEC-exp1,magBEC-exp2,magBEC-exp3,magBEC-exp4}. 
However, a theory~\cite{magBEC-theory} showed that 
if low-energy magnons of YIG are approximated by 
magnons of a ferromagnet without a sublattice structure, 
the magnon BEC is unstable due to the attractive interaction between magnons. 
Note first, that YIG is often treated as the ferromagnet 
for simplicity of analyses~\cite{YIG-review,YIG-Bauer}, 
second, in general, 
the attractive interaction between bosons destabilizes the BEC~\cite{FW,AGD}. 
Thus the stabilizing mechanism for the BEC of interacting magnons in a ferrimagnet 
remains unclear. 
To clarify it, 
we should understand the interaction effects in a ferrimagnet. 
In addition, 
we need to understand 
the essential effects of the differences between a ferrimagnet and the ferromagnet 
in order to understand the reason for the contradiction
between experiment~\cite{magBEC-exp1,magBEC-exp2,magBEC-exp3,magBEC-exp4} 
and theory~\cite{magBEC-theory}. 

In this Letter, 
we study the interaction effects on the magnon BEC in three magnets 
and propose a stabilizing mechanism. 
We use the Heisenberg Hamiltonian 
and consider a ferrimagnet and two ferromagnets. 
By using the Holstein-Primakoff transformation~\cite{HP,Oguchi,Nakamura}, 
we derive the kinetic energy and interaction for magnons. 
Then, we construct an effective theory to study 
the interaction effects on the magnon BEC 
in a similar way to the Bogoliubov theory~\cite{Bogoliubov,AGD} for Bose particles. 
By combining the results for the three magnets, 
we show that
the existence of a sublattice structure, 
not the difference in the spin alignment, 
is the key to 
the stabilizing mechanism for the BEC of interacting magnons. 
We also discuss 
the correspondence between our model and a more realistic model of YIG 
and several implications. 

We use 
the Heisenberg Hamiltonian as a minimal model for ferrimagnets and ferromagnets. 
It is given by
\begin{equation}
H=2\sum\limits_{\langle\boldi,\boldj\rangle}J_{\boldi\boldj}\boldS_{\boldi}\cdot\boldS_{\boldj},
\label{eq:H_spin}
\end{equation}
where $J_{\boldi\boldj}$ denotes the Heisenberg exchange energy
between spins at nearest-neighbor sites, 
and $\boldS_{\boldi}$ denotes the spin operator 
at site $\boldi$. 

\begin{figure*}
\includegraphics[width=114mm]{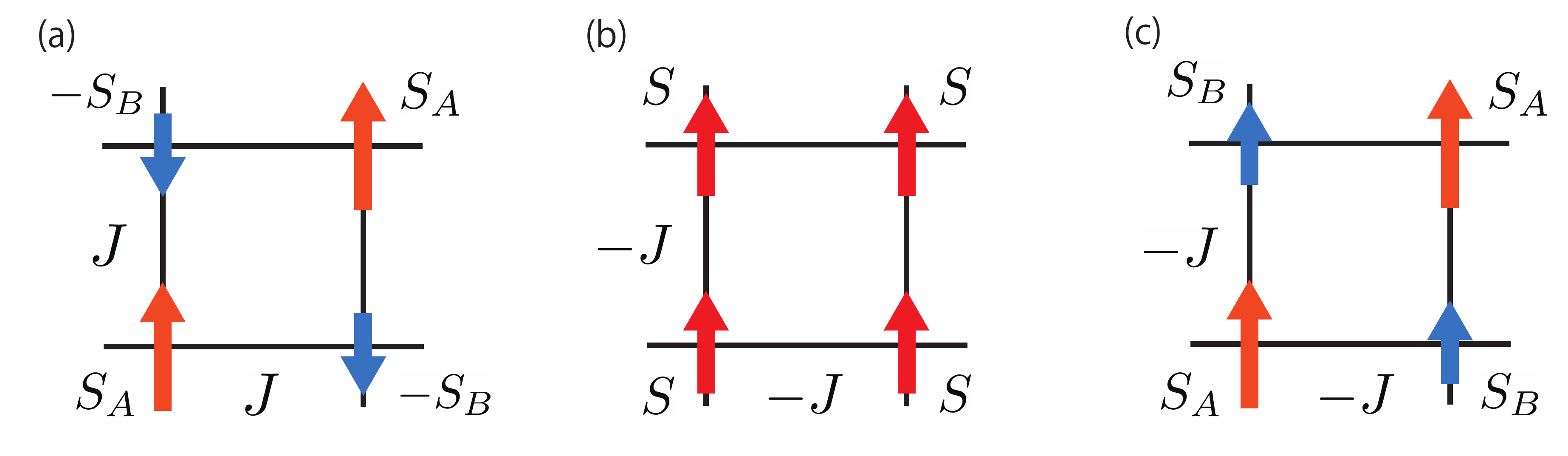}
\caption{\label{fig1} 
Spin alignments on a plane of the cubic lattice 
in the three cases of our model; 
panels (a), (b), and (c) correspond to the first, second, and third cases, 
respectively.  
The direction and length of an arrow represent the direction and size of an ordered spin. 
The ordered spins are ferrimagnetic in panel (a) and ferromagnetic in panels (b) and (c); 
sublattice degrees of freedom are present in panels (a) and (c) and absent in panel (b).}
\end{figure*}

We consider three cases. 
In the first case,
we put $J_{\boldi\boldj}=J$, $\langle\boldS_{\boldi}\rangle=S_{A}$ for $\boldi\in A$, 
and $\langle\boldS_{\boldi}\rangle=-S_{B}$ for $\boldi\in B$, 
where $A$ and $B$ denote $A$ and $B$ sublattices, respectively; 
each sublattice consists of $N/2$ sites.  
This case corresponds to 
a ferrimagnet with a two-sublattice structure [Fig. \ref{fig1}(a)]. 
In the second case, 
we put $J_{\boldi\boldj}=-J$ and $\langle\boldS_{\boldi}\rangle=S$ for all $\boldi$'s. 
In the third case, 
we put $J_{\boldi\boldj}=-J$, $\langle\boldS_{\boldi}\rangle=S_{A}$ for $\boldi\in A$, 
and $\langle\boldS_{\boldi}\rangle=S_{B}$ for $\boldi\in B$. 
The second and third cases correspond to 
ferromagnets without sublattice and with a two-sublattice structure, respectively 
[Figs. \ref{fig1}(b) and \ref{fig1}(c)]. 
As we will show below, 
by studying the BEC of interacting magnons in these three cases, 
we can clarify the stabilizing mechanism in a ferrimagnet 
and the key to resolving the contradiction in the magnon BEC of YIG.
(We will focus mainly on the sign of the effective interaction between magnons
and its effect on the stability of the magnon BEC.)

We begin with 
the first case of our model. 
We first derive the magnon Hamiltonian 
by using the Holstein-Primakoff transformation~\cite{HP,Oguchi,Nakamura}. 
After remarking on several properties in the BEC of noninteracting magnons, 
we construct the effective theory for the BEC of interacting magnons.  
By using this theory, we study the interaction effects in the ferrimagnet. 

The magnon Hamiltonian is obtained by applying 
the Holstein-Primakoff transformation to the spin Hamiltonian. 
In general, 
low-energy excitations in a magnet can be described well by magnons, 
bosonic quasiparticles~\cite{HP,Oguchi,Nakamura,Bloch,Dyson,Kubo,Harris,Manou}. 
The magnon operators and the spin operators are connected 
by the Holstein-Primakoff transformation~\cite{HP,Oguchi,Nakamura}. 
This transformation for our ferrimagnet is expressed as follows:
\begin{eqnarray}
S_{\boldi}^{z}=S_{A}-a_{\boldi}^{\dagger}a_{\boldi},\ 
S_{\boldi}^{-}=\sqrt{2S_{A}}a_{\boldi}^{\dagger}
\sqrt{1-\tfrac{a_{\boldi}^{\dagger}a_{\boldi}}{2S_{A}}},\label{eq:HP-A}\\
S_{\boldj}^{z}=-S_{B}+b_{\boldj}^{\dagger}b_{\boldj},  
S_{\boldj}^{+}=\sqrt{2S_{B}}b_{\boldj}^{\dagger}
\sqrt{1-\tfrac{b_{\boldj}^{\dagger}b_{\boldj}}{2S_{B}}},\label{eq:HP-B}
\end{eqnarray}
where $\boldi\in A$, $\boldj\in B$, 
$S_{\boldi}^{-}=S_{\boldi}^{x}-iS_{\boldi}^{y}=(S_{\boldi}^{+})^{\dagger}$, 
and $S_{\boldj}^{+}=S_{\boldj}^{x}+iS_{\boldj}^{y}=(S_{\boldj}^{-})^{\dagger}$; 
$a_{\boldi}$ and $a_{\boldi}^{\dagger}$ are the operators of magnons for the $A$ sublattice, 
and 
$b_{\boldj}$ and $b_{\boldj}^{\dagger}$ are those for the $B$ sublattice. 
A substitution of Eqs. (\ref{eq:HP-A}) and (\ref{eq:HP-B}) into Eq. (\ref{eq:H_spin}) 
gives the magnon Hamiltonian.  

In the magnon Hamiltonian, 
we consider the kinetic energy terms and the dominant terms
of the magnon-magnon interaction. 
This is because our aim is to clarify how the magnon-magnon interaction 
affects the magnon BEC, which is stabilized by the kinetic energy terms. 
Since the kinetic energy terms come from the quadratic terms of magnon operators 
and the dominant terms of the interaction 
come from part of the quartic terms~\cite{Oguchi,Nakamura}, 
our magnon Hamiltonian is given by $H_{\textrm{mag}}=H_{\textrm{non}}+H_{\textrm{int}}$~\cite{Supp}, where 
\begin{eqnarray}
H_{\textrm{non}}
=2\sum\limits_{\boldq}J(\boldzero)
(S_{B}a_{\boldq}^{\dagger}a_{\boldq}+S_{A}b_{\boldq}^{\dagger}b_{\boldq})\notag\\
+2\sum\limits_{\boldq}J(\boldq)
\sqrt{S_{A}S_{B}}
(a_{\boldq}b_{\boldq}+a_{\boldq}^{\dagger}b_{\boldq}^{\dagger}),\label{eq:H0-mag}
\end{eqnarray}
and 
\begin{eqnarray}
&H_{\textrm{int}}
=-\frac{2}{N}\sum\limits_{\boldq,\boldq^{\prime}}
[J(\boldzero)a_{\boldq}^{\dagger}a_{\boldq}b_{\boldq^{\prime}}^{\dagger}b_{\boldq^{\prime}}
+J(\boldq-\boldq^{\prime})a_{\boldq}^{\dagger}a_{\boldq^{\prime}}b_{\boldq}^{\dagger}b_{\boldq^{\prime}}\notag\\
&+\tfrac{J(\boldq)}{\sqrt{S_{A}S_{B}}}
(S_{A}a_{\boldq}b_{\boldq^{\prime}}^{\dagger}b_{\boldq}b_{\boldq^{\prime}}
+S_{B}b_{\boldq}a_{\boldq^{\prime}}^{\dagger}a_{\boldq^{\prime}}a_{\boldq})
]+(\textrm{H.c.}).\label{eq:Hint-mag}
\end{eqnarray}
We have used  
$a_{\boldi}=\sqrt{\tfrac{2}{N}}\sum_{\boldq}e^{i\boldq\cdot\boldi}a_{\boldq}$, 
$b_{\boldj}^{\dagger}=\sqrt{\tfrac{2}{N}}\sum_{\boldq}e^{i\boldq\cdot\boldj}b_{\boldq}^{\dagger}$, 
and 
$J(\boldq)=\sum_{\bolddelta}Je^{i\boldq\cdot\bolddelta}$ 
with $\bolddelta$, a vector to nearest neighbors. 

Before formulating the effective theory for the BEC of interacting magnons, 
we remark on several properties in the BEC of noninteracting magnons in our ferrimagnet. 
To see the properties, 
we diagonalize $H_{\textrm{non}}$ by using 
\begin{equation}
\left(
\begin{array}{c}
a_{\boldq} \\
b_{\boldq}^{\dagger}
\end{array}
\right)
=
\left(
\begin{array}{cc}
c_{\boldq} & -s_{\boldq} \\
-s_{\boldq} & c_{\boldq}
\end{array}
\right)
\left(
\begin{array}{c}
\alpha_{\boldq} \\
\beta_{\boldq}^{\dagger}
\end{array}
\right),\label{eq:Bogo-trans}
\end{equation}
where 
$c_{\boldq}\equiv\cosh \theta_{\boldq}$ and $s_{\boldq}\equiv\sinh \theta_{\boldq}$ satisfy  
$\tanh 2\theta_{\boldq}=\frac{2\sqrt{S_{A}S_{B}}J(\boldq)}{(S_{A}+S_{B})J(\boldzero)}$. 
After some algebra, we obtain 
\begin{equation}
H_{\textrm{non}}=\sum\limits_{\boldq}
\epsilon_{\alpha}(\boldq)\alpha_{\boldq}^{\dagger}\alpha_{\boldq}
+\sum\limits_{\boldq}
\epsilon_{\beta}(\boldq)\beta_{\boldq}^{\dagger}\beta_{\boldq},\label{eq:Hnon-mag}
\end{equation}
where 
$\epsilon_{\alpha}(\boldq)
=(S_{B}-S_{A})J(\boldzero)+\Delta \epsilon(\boldq)$ 
and $\epsilon_{\beta}(\boldq)
=(S_{A}-S_{B})J(\boldzero)+\Delta \epsilon(\boldq)$ 
with $\Delta \epsilon(\boldq)
=\sqrt{(S_{A}+S_{B})^{2}J(\boldzero)^{2}-4S_{A}S_{B}J(\boldq)^{2}}$; 
in Eq. (\ref{eq:Hnon-mag}) we have neglected the constant terms. 
Hereafter, we assume $S_{A}> S_{B}$; this does not lose generality. 
For $S_{A} > S_{B}$  
$\epsilon_{\alpha}(\boldzero)=0$ is the lowest energy. 
Thus many magnons occupy the $\boldq=0$ state of the $\alpha$ band 
in the BEC of noninteracting magnons in the ferrimagnet for $S_{A} > S_{B}$. 
In addition, 
the low-energy excitations from the condensed state 
are described by 
the $\alpha$-band magnons near $\boldq=\boldzero$. 

We now construct the effective theory for the BEC of interacting magnons. 
To construct it as simple as possible, 
we utilize the properties in the BEC of noninteracting magnons. 
As described above, in the ferrimagnet for $S_{A}>S_{B}$ 
the condensed state is the $\boldq=\boldzero$ state of the $\alpha$ band 
and the low-energy noncondensed states are the small-$\boldq$ states of the $\alpha$ band. 
Thus we can reduce $H_{\textrm{mag}}$ 
to an effective Hamiltonian $H_{\textrm{eff}}$, 
which consists of the kinetic energy term of the $\alpha$ band 
and the intraband terms of the magnon-magnon interaction for the $\alpha$ band; 
$H_{\textrm{eff}}$ is given by $H_{\textrm{eff}}=H_{0}+H^{\prime}$, 
where 
$H_{0}$ is the first term of Eq. (\ref{eq:Hnon-mag}), 
and 
$H^{\prime}$ is obtained by substituting 
Eq. (\ref{eq:Bogo-trans}) into Eq. (\ref{eq:Hint-mag})
and retaining the intraband terms. 
This $H_{\textrm{eff}}$ is sufficient 
for studying properties of the BEC of interacting magnons 
at temperatures lower than a Curie temperature, 
because the dominant excitations come from the small-$\boldq$ magnons 
in the $\alpha$ band 
and the interband terms may be negligible in comparison with the intraband terms. 
Then we can further simplify $H^{\prime}$.  
Since its main effects can be taken into account 
in the mean-field approximation, 
the leading term of $H^{\prime}$ is given by~\cite{Supp}
\begin{equation}
H^{\prime}=-\frac{4}{N}
\sum\limits_{\boldq,\boldq^{\prime}}\Gamma_{\alpha\alpha}(\boldq,\boldq^{\prime})n_{\boldq^{\prime}\alpha}
\alpha_{\boldq}^{\dagger}\alpha_{\boldq},\label{eq:Hint_MFA}
\end{equation}
where 
$\Gamma_{\alpha\alpha}(\boldq,\boldq^{\prime})
=J(\boldzero)(c_{\boldq}^{2}s_{\boldq^{\prime}}^{2}+c_{\boldq^{\prime}}^{2}s_{\boldq}^{2})
+2J(\boldq-\boldq^{\prime})c_{\boldq}s_{\boldq}c_{\boldq^{\prime}}s_{\boldq^{\prime}}
-\frac{J(\boldq)}{\sqrt{S_{A}S_{B}}}c_{\boldq}s_{\boldq}
(S_{A}s_{\boldq^{\prime}}^{2}+S_{B}c_{\boldq^{\prime}}^{2})
-\frac{J(\boldq^{\prime})}{\sqrt{S_{A}S_{B}}}c_{\boldq^{\prime}}s_{\boldq^{\prime}}
(S_{A}s_{\boldq}^{2}+S_{B}c_{\boldq}^{2})$,
and $n_{\boldq^{\prime}\alpha}=\langle\alpha_{\boldq^{\prime}}^{\dagger}\alpha_{\boldq^{\prime}}\rangle
=n[\epsilon_{\alpha}(\boldq^{\prime})]$ 
with the Bose distribution function $n(\epsilon)$.  
By combining Eq. (\ref{eq:Hint_MFA}) with 
$H_{0}=\sum_{\boldq}\epsilon_{\alpha}(\boldq)\alpha_{\boldq}^{\dagger}\alpha_{\boldq}$, 
we obtain
\begin{equation}
H_{\textrm{eff}}
=\sum\limits_{\boldq}\epsilon_{\alpha}^{\ast}(\boldq)\alpha_{\boldq}^{\dagger}\alpha_{\boldq},\label{eq:remark}
\end{equation}
with $\epsilon_{\alpha}^{\ast}(\boldq)=\epsilon_{\alpha}(\boldq)
-\frac{4}{N}\sum_{\boldq^{\prime}}\Gamma_{\alpha\alpha}(\boldq,\boldq^{\prime})n_{\boldq^{\prime}\alpha}$. 

By using the theory described by $H_{\textrm{eff}}$, 
we study the interaction effects on the stability of the magnon BEC. 
Since the magnon energy should be nonnegative, 
the magnon BEC remains stable even for interacting magnons 
as long as $\epsilon_{\alpha}^{\ast}(\boldzero)$ is the lowest energy. 
This is realized if $H^{\prime}$ is the repulsive interaction. 
If $H^{\prime}$ is the attractive interaction, 
the magnon BEC becomes unstable. 
Thus we need to analyze the sign of $\Gamma_{\alpha\alpha}(\boldq,\boldq^{\prime})$ 
in Eq. (\ref{eq:Hint_MFA}). 
Since the dominant low-energy excitations are described by the $\alpha$-band magnons 
near $\boldq=\boldzero$, 
we estimate $\Gamma_{\alpha\alpha}(\boldq,\boldq^{\prime})$ in Eq. (\ref{eq:Hint_MFA}) 
in the long-wavelength limits $|\boldq|,|\boldq^{\prime}|\rightarrow 0$. 
For a concrete simple example 
we perform this estimation in a three-dimensional case on the cubic lattice. 
By expressing $J(\boldq)$ in a Taylor series around $|\boldq|=0$ 
and retaining the leading correction, 
we get $J(\boldq)\approx J(\boldzero)[1-\frac{q^{2}}{6}]$. 
Then, by using this expression and performing some calculations~\cite{Supp}, 
we obtain the expression of $\Gamma_{\alpha\alpha}(\boldq,\boldq^{\prime})$ including 
the leading correction in the long-wavelength limits. 
The derived expression is 
\begin{equation}
\Gamma_{\alpha\alpha}(\boldq,\boldq^{\prime})\approx -\frac{2}{9}
J(\boldzero)q^{2}q^{\prime 2}\frac{(S_{A}S_{B})^{2}}{(S_{A}-S_{B})^{4}}.
\label{eq:Gam-approx}
\end{equation}
The combination of Eqs. (\ref{eq:Gam-approx}) and (\ref{eq:Hint_MFA}) shows that 
the leading term of the magnon-magnon interaction is repulsive. 
Thus 
the magnon BEC remains stable in the ferrimagnet even with the magnon-magnon interaction.

The above 
result differs from the stability of the magnon BEC 
in the ferromagnet without a sublattice structure. 
This can be seen by applying a similar theory to the second case of our model 
and comparing the result with the above result. 
The Holstein-Primakoff transformation in the ferromagnet without a sublattice structure 
is expressed as 
$S_{\boldi}^{z}=S-c_{\boldi}^{\dagger}c_{\boldi}$, 
$S_{\boldi}^{-}=c_{\boldi}^{\dagger}\sqrt{2S-c_{\boldi}^{\dagger}c_{\boldi}}$, 
and $S_{\boldi}^{+}=(S_{\boldi}^{-})^{\dagger}$ for all $\boldi$'s; 
$c_{\boldi}$ and $c_{\boldi}^{\dagger}$ are the magnon operators. 
By using this transformation and the Fourier transformations 
of the magnon operators, such as 
$c_{\boldi}=\frac{1}{\sqrt{N}}\sum_{\boldq}e^{i\boldq\cdot \boldi}c_{\boldq}$, 
we obtain the magnon Hamiltonian $H_{\textrm{mag}}
=H_{\textrm{non}}+H_{\textrm{int}}$, 
where 
$H_{\textrm{non}}=\sum_{\boldq}\epsilon(\boldq)c_{\boldq}^{\dagger}c_{\boldq}$ 
with $\epsilon(\boldq)=2S[J(\boldzero)-J(\boldq)]$ 
and 
$H_{\textrm{int}}=-\frac{1}{2N}\sum_{\boldq,\boldq^{\prime}}
[J(\boldzero)c^{\dagger}_{\boldq}c_{\boldq}c_{\boldq^{\prime}}^{\dagger}c_{\boldq^{\prime}}
+J(\boldq-\boldq^{\prime})c_{\boldq}^{\dagger}c_{\boldq^{\prime}}c_{\boldq^{\prime}}^{\dagger}c_{\boldq}
-2J(\boldq)c_{\boldq^{\prime}}^{\dagger}c_{\boldq}c_{\boldq^{\prime}}^{\dagger}c_{\boldq}]
+(\textrm{H.c.})$. 
Then, by applying the mean-field approximation to $H_{\textrm{int}}$, 
the leading term of the magnon-magnon interaction is reduced to 
$H^{\prime}=-\frac{2}{N}\sum_{\boldq,\boldq^{\prime}}
\Gamma(\boldq,\boldq^{\prime})n_{\boldq^{\prime}}c_{\boldq}^{\dagger}c_{\boldq}$, 
where $\Gamma(\boldq,\boldq^{\prime})
=J(\boldzero)+J(\boldq-\boldq^{\prime})-J(\boldq)-J(\boldq^{\prime})$ 
and $n_{\boldq^{\prime}}\equiv n[\epsilon(\boldq^{\prime})]$. 
Since $\Gamma(\boldq,\boldq^{\prime})\ge 0$, 
the magnon-magnon interaction becomes attractive. 
Thus the BEC of interacting magnons becomes unstable 
in the ferromagnet without a sublattice structure. 

In order 
to understand the key to causing the above difference, 
we study the stability of the BEC of interacting magnons in the third case of our model. 
As we can see from Fig. \ref{fig1}, 
the difference between the third and first cases is about the spin alignment, 
and the difference between the third and second cases is about the sublattice structure. 
Thus, by comparing the result in the third case 
with the result in the first or second case, 
we can deduce which of the two, the differences 
in the spin alignment and in the sublattice structure, causes 
the difference in the stability of the BEC of interacting magnons. 

The stability in the third case can be studied in a similar way to that in the first case. 
In the third case, 
the Holstein-Primakoff transformation of $\boldS_{\boldi}$ for $\boldi\in A$ 
is the same as Eq. (\ref{eq:HP-A}), 
whereas that of $\boldS_{\boldj}$ for $\boldj\in B$ 
is given by 
$S_{\boldj}^{z}=S_{B}-b_{\boldj}^{\dagger}b_{\boldj}$,
$S_{\boldj}^{-}=\sqrt{2S_{B}}b_{\boldj}^{\dagger}
\sqrt{1-(b_{\boldj}^{\dagger}b_{\boldj}/2S_{B})}$, 
and $S_{\boldj}^{+}=(S_{\boldj}^{-})^{\dagger}$; 
this difference arises from the different alignment of the spins belonging to the $B$ sublattice. 
In a similar way to the first case, 
we obtain 
the magnon Hamiltonian 
$H_{\textrm{mag}}=H_{\textrm{non}}+H_{\textrm{int}}$, where 
$H_{\textrm{non}}$ and $H_{\textrm{int}}$ are given by
\begin{eqnarray}
H_{\textrm{non}}
=2\sum\limits_{\boldq}J(\boldzero)
(S_{B}a_{\boldq}^{\dagger}a_{\boldq}+S_{A}b_{\boldq}^{\dagger}b_{\boldq})\notag\\
-2\sum\limits_{\boldq}J(\boldq)
\sqrt{S_{A}S_{B}}
(a_{\boldq}b_{\boldq}^{\dagger}+a_{\boldq}^{\dagger}b_{\boldq}),
\end{eqnarray}
and 
\begin{eqnarray}
& &H_{\textrm{int}}
=-\frac{2}{N}\sum\limits_{\boldq,\boldq^{\prime}}
[J(\boldzero)a_{\boldq}^{\dagger}a_{\boldq}b_{\boldq^{\prime}}^{\dagger}b_{\boldq^{\prime}}
+J(\boldq-\boldq^{\prime})a_{\boldq}^{\dagger}a_{\boldq^{\prime}}b_{\boldq^{\prime}}^{\dagger}b_{\boldq}\notag\\
& &
-\tfrac{J(\boldq)}{\sqrt{S_{A}S_{B}}}
(S_{A}a^{\dagger}_{\boldq}b_{\boldq^{\prime}}^{\dagger}b_{\boldq^{\prime}}b_{\boldq}
+S_{B}b_{\boldq}a_{\boldq}^{\dagger}a_{\boldq^{\prime}}^{\dagger}a_{\boldq^{\prime}})
]+(\textrm{H.c.}),
\end{eqnarray}
respectively,
with $a_{\boldi}=\sqrt{\tfrac{2}{N}}\sum_{\boldq}e^{i\boldq\cdot\boldi}a_{\boldq}$ and 
$b_{\boldj}=\sqrt{\tfrac{2}{N}}\sum_{\boldq}e^{i\boldq\cdot\boldj}b_{\boldq}$.  
In addition, $H_{\textrm{non}}$ can be diagonalized 
by using 
$a_{\boldq}=c_{\boldq}\alpha_{\boldq}-s_{\boldq}\beta_{\boldq}$ 
and $b_{\boldq}=-s_{\boldq}\alpha_{\boldq}+c_{\boldq}\beta_{\boldq}$, 
where $c_{\boldq}\equiv\cosh \theta_{\boldq}$ and $s_{\boldq}\equiv\sinh \theta_{\boldq}$ 
satisfy 
$\tanh 2\theta_{\boldq}=-\frac{2\sqrt{S_{A}S_{B}}J(\boldq)}{(S_{A}+S_{B})J(\boldzero)}$.
The diagonalized $H_{\textrm{non}}$ is 
$H_{\textrm{non}}=\sum_{\boldq}
[\epsilon_{\alpha}(\boldq)\alpha_{\boldq}^{\dagger}\alpha_{\boldq}
+\epsilon_{\beta}(\boldq)\beta_{\boldq}^{\dagger}\beta_{\boldq}]$ 
with $\epsilon_{\alpha}(\boldq)$ and $\epsilon_{\beta}(\boldq)$, 
which are the same as those in the first case. 
Thus, the ferromagnet and ferrimagnet with the two-sublattice structure 
have the same properties of the BEC of noninteracting magnons. 
Then 
we can construct the effective theory for the BEC of interacting magnons 
in the third case in a similar way. 
For $S_{A}>S_{B}$, in the third case, 
the BEC of interacting magnons can be effectively described 
by $H_{\textrm{eff}}
=\sum_{\boldq}\epsilon_{\alpha}^{\ast}(\boldq)\alpha_{\boldq}^{\dagger}\alpha_{\boldq}$ 
with $\epsilon_{\alpha}^{\ast}(\boldq)=\epsilon_{\alpha}(\boldq)
-\frac{4}{N}\sum_{\boldq^{\prime}}\tilde{\Gamma}_{\alpha\alpha}(\boldq,\boldq^{\prime})n_{\boldq^{\prime}\alpha}$, 
where 
$\tilde{\Gamma}_{\alpha\alpha}(\boldq,\boldq^{\prime})=
J(\boldzero)(c_{\boldq}^{2}s_{\boldq^{\prime}}^{2}+c_{\boldq^{\prime}}^{2}s_{\boldq}^{2})
+2J(\boldq-\boldq^{\prime})c_{\boldq}s_{\boldq}c_{\boldq^{\prime}}s_{\boldq^{\prime}}
+\frac{J(\boldq)}{\sqrt{S_{A}S_{B}}}c_{\boldq}s_{\boldq}
(S_{A}s_{\boldq^{\prime}}^{2}+S_{B}c_{\boldq^{\prime}}^{2})
+\frac{J(\boldq^{\prime})}{\sqrt{S_{A}S_{B}}}c_{\boldq^{\prime}}s_{\boldq^{\prime}}
(S_{A}s_{\boldq}^{2}+S_{B}c_{\boldq}^{2})$. 
By estimating $\tilde{\Gamma}_{\alpha\alpha}(\boldq,\boldq^{\prime})$ 
in the long-wavelength limits in a similar way, 
we obtain 
$\tilde{\Gamma}_{\alpha\alpha}(\boldq,\boldq^{\prime})\approx -\frac{2}{9}
J(\boldzero)q^{2}q^{\prime 2}\frac{(S_{A}S_{B})^{2}}{(S_{A}-S_{B})^{4}}$. 
Thus the BEC of interacting magnons is stable in the ferromagnet 
with the two-sublattice structure. 

Combining the results in the three cases, 
we find that 
the difference between the interaction effects in the ferrimagnet 
and in the ferromagnet without a sublattice structure arises 
not from the difference in the spin alignment, 
but from the difference in the sublattice structure. 
This can resolve the contradiction 
between experiment~\cite{magBEC-exp1,magBEC-exp2,magBEC-exp3,magBEC-exp4} 
and theory~\cite{magBEC-theory} 
because that theory uses a ferromagnet without a sublattice structure. 
This also suggests that 
the existence of a sublattice structure 
is the key to stabilizing the BEC of interacting magnons in ferrimagnets and ferromagnets. 
One possible experiment to test our mechanism
is to measure the stability of the magnon BEC
in ferromagnets without and with a sublattice structure; 
a sublatttice structure, such as that shown in Fig. \ref{fig1}(c),
can be realized, for example, by using two different magnetic ions. 

We remark on the role of sublattice degrees of freedom. 
As shown above, 
the magnon BEC remains stable even in the presence of the magnon-magnon interaction 
as long as a magnet has the sublattice degrees of freedom. 
This remarkable property can hardly be expected from 
the properties of noninteracting magnons 
because in all the three cases, 
the low-energy properties can be described 
by a single magnon band. 
The magnon-magnon interaction becomes repulsive 
only in the presence of the sublattice degrees of freedom 
because the magnons in different sublattices give 
the different contributions to the intraband interaction for a single magnon band; 
the different contributions arise from the different coefficients 
in the Bogoliubov transformation [e.g., see Eq. (\ref{eq:Bogo-trans})]. 

Next we discuss the correspondence between our model 
and a model derived in the first-principles study in YIG~\cite{YIG-1stPrinciple}. 
The latter is more complicated than our model 
because the magnetic primitive cell of YIG has $20$ Fe moments~\cite{Harris2} 
and its spin Hamiltonian consists of 
the Heisenberg exchange interactions for three nearest-neighbor pairs 
and six next-nearest-neighbor pairs~\cite{YIG-1stPrinciple}. 
Note first, that 
all of the Fe ions are categorized into 
Fe$^{\textrm{O}}$ and Fe$^{\textrm{T}}$ ions, 
Fe ions surrounded by an octahedron and a tetrahedron of O ions, respectively, 
and second, that YIG is a ferrimagnet due to 
the antiparallel spin alignments of the Fe$^{\textrm{O}}$ and Fe$^{\textrm{T}}$ ions 
and the $2:3$ ratio of the Fe$^{\textrm{O}}$ and Fe$^{\textrm{T}}$ ions 
in the unit cell~\cite{YIG-Ferri}. 
Although our model does not take into account all of the complex properties of YIG, 
our model can be regarded as a minimal model to 
study the stability of the BEC of interacting magnons in YIG. 
This is because of the following three facts: 
First, 
the largest term in the spin Hamiltonian of YIG is the antiferromagnetic nearest-neighbor 
Heisenberg exchange interaction between the Fe$^{\textrm{O}}$ and Fe$^{\textrm{T}}$ ions 
and the others are at least an order of magnitude smaller. 
Second, 
the low-energy magnons of YIG can be described by
a single magnon band around $\boldq=\boldzero$. 
Third, 
the main effect of the terms neglected in our theory is 
to modify the value of $\Gamma_{\alpha\alpha}(\boldq,\boldq^{\prime})$ 
in Eq. (\ref{eq:Hint_MFA}). 
Since this modification may be quantitative, 
our mechanism can qualitatively explain why the magnon BEC is stabilized in YIG. 

Our work has several implications. 
First, 
our results suggest that 
a ferromagnet without a sublattice structure is inappropriate for describing the properties 
of interacting magnons in ferrimagnets, such as YIG. 
This suggestion will be useful for future studies towards 
a comprehensive understanding of magnon physics 
and spintronics using magnons in YIG. 
Furthermore, 
it may be necessary to reconsider some results of YIG 
if the results are deduced by using a ferromagnet
without a sublattice structure, 
in particular, the results depend on the sign of the magnon-magnon interaction. 
Our theoretical framework can then be used to study the BEC of interacting magnons 
in other magnets as long as the low-energy magnons can be described by 
a single magnon band. 
For the magnets whose low-energy magnons have degeneracy, 
an extension of this framework enables us to study the BEC of interacting magnons. 
Thus our theory may provide a starting point for understanding 
properties of the BEC of interacting magnons in various magnets.

In summary, 
we have studied the stability of the BEC of interacting magnons 
in a ferrimagnet and ferromagnets, 
and we proposed the stabilizing mechanism. 
By adopting the Holstein-Primakoff transformation to the Heisenberg Hamiltonian, 
we have derived the magnon Hamiltonian, which consists of 
the kinetic energy terms and the dominant terms of the magnon-magnon interaction. 
We then construct the effective theory for the BEC of interacting magnons 
by utilizing the properties for noninteracting magnons and the mean-field approximation. 
From the analyses using this theory, 
we have deduced that 
in the ferrimagnet and ferromagnet with the sublattice structure 
the magnon BEC remains stable even in the presence of the magnon-magnon interaction, 
whereas it becomes unstable in the ferromagnet without a sublattice. 
This result shows that 
the existence of a sublattice structure is the key to stabilizing the BEC of interaction magnons, 
whereas the difference in the spin alignments is irrelevant. 
In addition, 
this result is consistent with 
the experimental results~\cite{magBEC-exp1,magBEC-exp2,magBEC-exp3,magBEC-exp4} of YIG 
and the theoretical result~\cite{magBEC-theory} of a ferromagnet without a sublattice structure.  


\end{document}